\documentclass[prb,preprint]{revtex4-2} 
\usepackage{amsmath}  % needed for \tfrac, \bmatrix, etc.
\usepackage{amsfonts} % needed for bold Greek, Fraktur, and blackboard bold
\usepackage{graphicx} % needed for figures

\usepackage{enumitem}

\def\ben{\begin{equation}}
\def\een{\end{equation}}

  \let\n=\nu

\let\C=\Chi

 \def\bd{\begin{document}} \def\ed{\end{document}}
\def\ds{\documentstyle} \let\fr=\frac \let\bl=\bigl \let\br=\bigr
\let\Br=\Bigr \let\Bl=\Bigl
\let\bm=\bibitem
\let\na=\nabla
\let\pa=\partial \let\ov=\overline
\newcommand{\be}{\begin{equation}}
\newcommand{\ee}{\end{equation}}
\def\ba{\begin{array}}
\def\ea{\end{array}}
\def\ft#1#2{{\textstyle{{\scriptstyle #1}\over {\scriptstyle #2}}}}
\def\fft#1#2{{#1 \over #2}}
\def\del{\partial}
\def\vp{\varphi}
\def\sst#1{{\scriptscriptstyle #1}}
\def\oneone{\rlap 1\mkern4mu{\rm l}}
\def\td{\tilde}
\def\wtd{\widetilde}
\def\ie{\rm i.e.\ }
\def\dalemb#1#2{{\vbox{\hrule height .#2pt
        \hbox{\vrule width.#2pt height#1pt \kern#1pt
                \vrule width.#2pt}
        \hrule height.#2pt}}}
\def\square{\mathord{\dalemb{6.8}{7}\hbox{\hskip1pt}}}
\newcommand{\ho}[1]{$\, ^{#1}$}
\newcommand{\hoch}[1]{$\, ^{#1}$}
\newcommand{\bea}{\begin{eqnarray}}
\newcommand{\eea}{\end{eqnarray}}
\newcommand{\ra}{\rightarrow}
\newcommand{\lra}{\longrightarrow}
\newcommand{\Lra}{\Leftrightarrow}
\newcommand{\bp}{\tilde \beta^\prime}
\newcommand{\tr}{{\rm tr} }
\newcommand{\Tr}{{\rm Tr} }
\def\0{{\sst{(0)}}}
\def\1{{\sst{(1)}}}
\def\2{{\sst{(2)}}}
\def\3{{\sst{(3)}}}
\def\4{{\sst{(4)}}}
\def\5{{\sst{(5)}}}
\def\6{{\sst{(6)}}}
\def\7{{\sst{(7)}}}
\def\8{{\sst{(8)}}}
\def\n{{\sst{(n)}}}
\def\cA{{{\cal A}}}
\def\cB{{{\cal B}}}
\def\cF{{{\cal F}}}
\def\cH{{{\cal H}}}
\def\tV{\widetilde V}
\def\tW{\widetilde W}
\def\tH{\widetilde H}
\def\tE{\widetilde E}
\def\tF{\widetilde F}
\def\tA{\widetilde A}
\def\im{{i}}
\def\tY{{{\wtd Y}}}
\def\ep{{\epsilon}}
\def\vep{{\varepsilon}}
\def\R{\rlap{\rm I}\mkern3mu{\rm R}}
\def\bD{{{\bar D}}}

\def\R{\rlap{\rm I}\mkern3mu{\rm R}}
\def\bD{{{\bar D}}}
\def\R{{{\Bbb R}}}
\def\C{{{\Bbb C}}}
\def\H{{{\Bbb H}}}
\def\CP{{{\Bbb C}{\Bbb P}}}
\def\RP{{{\Bbb R}{\Bbb P}}}
\def\Z{{{\Bbb Z}}}
\def\bA{{{\Bbb A}}}
\def\bB{{{\Bbb B}}}
\def\bC{{{\Bbb C}}}
\def\bD{{{\Bbb D}}}
\def\bE{{{\Bbb E}}}
\def\bZ{{{\Bbb Z}}}
\def\Re{{{\frak{Re}}}}
\def\Im{{{\frak{Im}}}}
\def\cosec{{\,\hbox{cosec}\,}}
\def\Gm{{\Gamma_{\!\! -}}}
\def\Gp{{\Gamma_{\!\! +}}}
\def\stan{{standard }}
\def\nonstan{{supernumerary }}

\newcommand{\auth}{Zhiwei Chong}
\begin{document}

% Be sure to use the \title, \author, \affiliation, and \abstract macros
% to format your title page.  Don't use lower-level macros to  manually
% adjust the fonts and centering.

%\title{A Close Look at Inelastic Collisions}
% In a long title you can use \\ to force a line break at a certain location.
\vspace{10pt}

\begin{center}

{\large {\bf A Qualitative Analysis to Simple Harmonic Motion}}\\
\vspace{5pt}
%\auth\\

%\vspace{10pt}
%{\it International Division, Experimental School Affiliated with Zhuhai No.1 High School, Zhuhai, Guangdong, China}\\

%\vspace{10pt}{ \footnote{runnerwei@qq.com} \it Zhuhai No.1 High School, Zhuhai, Guangdong, China}

\end{center}

%When submitting the manuscript for review, do not include the author's name or institution
%\author{Daniel V. Schroeder}
%\email{dschroeder@weber.edu} % optional
%\altaffiliation[permanent address: ]{101 Main Street, Anytown, USA} % optional second address
% If there were a second author at the same address, we would put another 
% \author{} statement here.  Don't combine multiple authors in a single
% \author statement.
%\affiliation{Department of Physics, Weber State University, Ogden, UT 84408-2508}
% Please provide a full mailing address here.

%\author{David P. Jackson}
%\email{ajp@dickinson.edu}
%\affiliation{Department of Physics, Dickinson College, Carlisle, PA 17013}

% See the REVTeX documentation for more examples of author and affiliation lists.

\author{Zhiwei Chong}
\email{chong.zhiwei@gmail.com}
%\altaffiliation[permanent address: ]{101 Main Street, Anytown, USA} % optional second address
% If there were a second author at the same address, we would put another 
% \author{} statement here.  Don't combine multiple authors in a single
% \author statement.
\affiliation{International Division, Experimental School Affiliated with Zhuhai No.1 High School,\\ Zhuhai, Guangdong, China}
\date{\today}
\begin{abstract}
This paper proposes a qualitative analysis to the simple harmonic motion for students who are not mathematically well-prepared.
It uses the variation in speed and acceleration to sketch the velocity-time curve. 
The curve appears to be sinusoidal, whose shape is largely determined by the fact that the gradient of a velocity-time graph is acceleration. 
This approach  cannot determine the exact value of period and claim the curve to be sinusoidal, however, it complements rigorous mathematical analysis in that it is full of physical reasoning that the mathematical approaches lack~of.
\end{abstract}
% AJP requires an abstract for all regular article submissions.
% Abstracts are optional for submissions to the "Notes and Discussions" section.

\maketitle % title page is now complete
\newpage

Simple Harmonic Motion (henceforth SHM) is taught at various levels depending on students' maturity in mathematics \cite{coderive}. 
%If the course is based on algebra, then SHM is invariably introduced with the aid of uniform circular motion. 
%If the course is based on calculus, then the differential equation is presented. 
%However, there are three ways to obtain its solution. 
%Firstly, the solution is presented and students are asked to verify its validity
%Secondly, the solution is obtained by integrating twice.
%Lastly, the solution is obtained by solving the ordinary differential equation.
Without delving into much mathematics, this paper \textit{qualitatively} analyzes SHM based on elementary knowledge in kinematics, Newton's second law, Hooke's law, and mechanical energy conservation, which are usually exposed to students before the introduction of SHM.
%The fact from kinematics, that is, the gradient on a velocity-time curve is acceleration, is key to sketching the velocity-time curve.
The results from the analysis are used to sketch the velocity-time curve, and the shape of the curve is largely determined by the fact that the gradient of a velocity-time curve is acceleration.
Even though the obtained curve cannot be claimed to be sinusoidal, however, whoever sees it would strongly suspect that it~is.
The value of this approach is that key features of the seemingly complicated motion can be revealed to students who are not mathematically well prepared. 
Moreover, students are exposed to intensive physical reasoning rather than dry mathematical manipulations even though they are mathematically prepared.

The qualitative approach in this paper and rigorous mathematical approaches, either referring to uniform circular motion or solving differential equation \cite{giancoli, serway}, actually complement with each other. 
%On one hand, the qualitative approach cannot obtain the exact value of the period for SHM, and the sketched curve cannot be claimed to be sinusoidal. They have to be derived from rigorous mathematical approaches mentioned above.
On the one hand, rigorous approaches give the exact value of the period and confirm that the velocity-time curve is indeed sinusoidal, which the qualitative approach is unable to provide.
On the other hand, the qualitative approach is full of rich physics content that the mathematical approaches severely lack of. 
In particular, one of the two rigorous approaches, which refers to uniform circular motion, is criticized as nonphysical, arbitrary, artificial, imaginary, and even confusing to students \cite{novel, noncalculus, treatment, introducing}.

\section{Problem and Background Knowledge}\label{background}
A mass $m$ is hooked with one end of a massless spring with stiffness $k$. 
The other end of the spring is fixed on a wall.
The mass lies on a frictionless floor as illustrated in Fig. 1.
\begin{figure}[h!]
\centering
\includegraphics[width=4.0in]{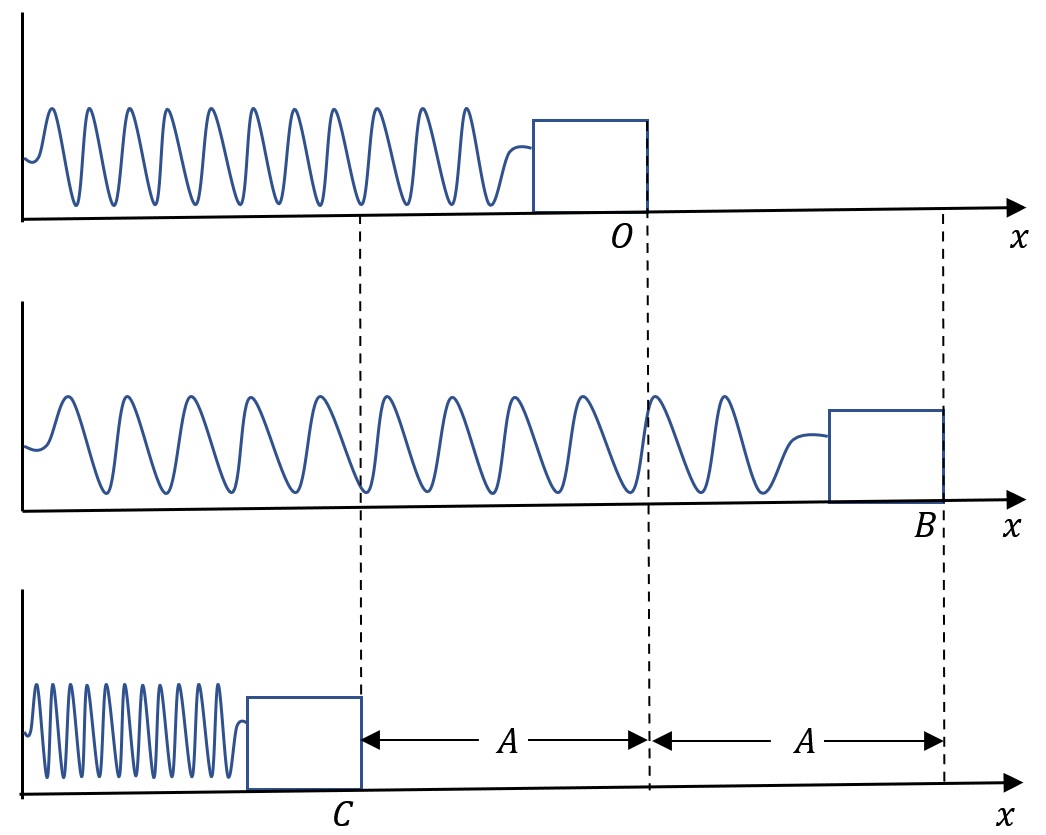}
%\hspace{1cm}。jpg
%\label{setup}
\caption{Setup of the Problem.}
\centering
\end{figure}
The equilibrium position is chosen as origin of the $x$-axis, and it is denoted as $O$. 
The mass is pulled to the right to point $B$ by distance $A$ 
(usually $A$ is reserved as the notation for amplitude).
Then it is released from rest. 
The mass moves back to equilibrium $O$. It won't stop at $O$ but keeps moving to the left. The point when it is momentarily at rest before moving right back to $O$ is denoted as $C$. 
The task is to sketch the velocity-time curve.

Before introducing SHM, the following facts were made clear to students.
They are the basis for the qualitative analysis and are listed in the order of the time students learned them.
%\renewcommand{\theenumi}{Fact\arabic{enumi}}
%\begin{enumerate}[label=Fact \arabic*:, wide=0pt, leftmargin=*,labelindent=\parindent, noitemsep]
\begin{enumerate}[label=Fact \arabic*:, wide=0pt, leftmargin=*,labelindent=\parindent, noitemsep]
%	\item  The gradient of a displacement-time graph is velocity.
	\item  The gradient of a velocity-time curve is acceleration.
	\item  Newton's second law: the acceleration is proportional to the net force.
	\item  Hooke's law: the elastic force is proportional to its deformation.
	\item  Mechanical energy is conserved: the sum of elastic potential energy and kinetic energy is equal to the initial elastic potential energy, that is, $\fft12\,k\,A^2$. 
\end{enumerate}

\section{Qualitative Analysis}\label{analysis}
%\noindent 
With the above listed facts at hand, a few qualitative results are obtained for SHM. 
They are listed below followed by explanations leading to them. 
In Sec. \ref{velocity-time}, they will be used to sketch the velocity-time curve.
\begin{enumerate}[label=Result \arabic*:, wide=0pt, leftmargin=*,labelindent=\parindent, noitemsep]
	\item The motion is periodic.
	\item The separation between equilibrium $O$ and $B$ is the same as that between $O$ and $C$, that is, $|OC|\,=\,|BO|\,=\,A$.
	\item For  two symmetric points with respect to the equilibrium $O$, the mass passes them with the same speed.
	\item It takes the same amount of time to cover each of the four segments: $B				\rightarrow O$, $O\rightarrow~C$, $C\rightarrow~O$, and $O\rightarrow B$. In other words, the time spent in each segment is $\fft14\,T$.
	\item As the mass moves from $B$ to $O$, its speed increases at a decreasing rate, and the direction is negative.
	\item As the mass moves from $O$ to $C$, its speed decreases at an increasing rate, and the direction is still negative.
	\item As the mass moves from $C$ back to $O$, its speed increases at an increasing rate, and the direction is positive.
	\item As the mass moves from $O$ back to $B$, its speed decreases at an increasing rate, and the direction is positive.
\end{enumerate}

As for Result 1, it is not hard to convince students that the motion is periodic, which enables us to focus curve sketching within only one period, even though its exact value is unknown with the qualitative analysis here.

Result 2 comes from Fact 4, that is, mechanical energy conservation. 
The kinetic energy is zero at both $B$ and $C$. 
Thereby, the elastic potential energy is the same, so is the extension for the spring.
Result~3 also comes from Fact~4.
Equal distance from the equilibrium means equal elastic potential energy. 
Thereby, the kinetic energy at these two points is the same, so is the speed.
Result 3 is used to obtain Result 4.

It takes a while to explain Result 4. 
I here only show that it takes the same amount of time to move from $B$ to $O$ as that from $C$ to $O$. 
The rest is left as an exercise for the reader. 
According to Result 3, for any two points that are symmetric with respect to the equilibrium $O$, the mass travels with the same speed. 
The only difference is direction; the mass moves in opposite directions at the two symmetric points. 
However, the time to cover $BO$ or $CO$ does not depend on moving direction.  
Therefore, the time to cover $BO$ is the same as~$CO$. 
Result 4 enables us to divide the period into 4 equal time intervals.

The analyses for Results 5-8 are similar, thereby, only the reasoning for 
Result~5 is given. 
As the mass moves from $B$ to $O$, its separation from the equilibrium $O$ decreases. According to Fact~4, its kinetic energy increases, so is its speed. On the other hand, according to Fact~3, the magnitude of the elastic force on it decreases, so is its magnitude of acceleration by Fact~2, or Newton's second law.

Results~5-8 are important for curve sketching in Sec. \ref{velocity-time}.
They, together with Fact~1, determine the shape of the velocity-time curve in crucial ways. 
To be specific, they determine the monotonicity, concavity, and convexity of the curve.

\section{Velocity-time Curve}\label{velocity-time}
Before details of the curve are sketched, the following are listed as preparation without explanation.
\begin{enumerate}[noitemsep]
	\item The curve is sketched within one period only by Result~1.
	\item The period $T$ is divided into 4 equal intervals by Result~4.
	\item At $t=0, \fft12\,T$, and $T$ the velocity is zero. 
	\item At $t=\fft14\,T, \fft34\,T$, the velocities are $-v_m$ and~$+v_m$, respectively, where $v_m$ is the maximal speed and $v_m\,=\,A\sqrt{k/m}$.
	\item For $t\in (0, \fft12\,T)$, the velocity is negative; the curve is below the time axis.
	\item For $t\in~(\fft12\,T, T)$, the velocity is positive; the curve is above the time axis.
	%\item At $t=\fft14\,T$ and $t=\fft34\,T$, the velocities are $-v_m$ and $+v_m$, respectively.
\end{enumerate}
More details about the curve are discussed for each of the 4 time intervals.
Please refer to Fig.~2 when reading the following subsections.

\subsection{ For $t\in[0, \fft14\,T]$ or $B\rightarrow O$:}
\begin{enumerate}[noitemsep]
	%\item At $t=0$ or at point $B$, it is at rest or $v=0$.
	\item The mass moves to the left or the negative direction; the curve is below the time axis.
	\item Its speed increases by Result~5; the curve becomes more and more negative.
	\item The magnitude of acceleration decreases by Result~5; the tangent to the curve becomes more and more shallow. Note the tangent at $P$ is shallower than that at $Q$ in Fig. 2.
	\item At $t=\fft14\,T$ or at point~$O$, the speed achieves its maximum $v_m$ and its acceleration is zero; the tangent to the curve is horizontal.
\end{enumerate}

\begin{figure}[h!]
\centering
\includegraphics[width=5.0in]{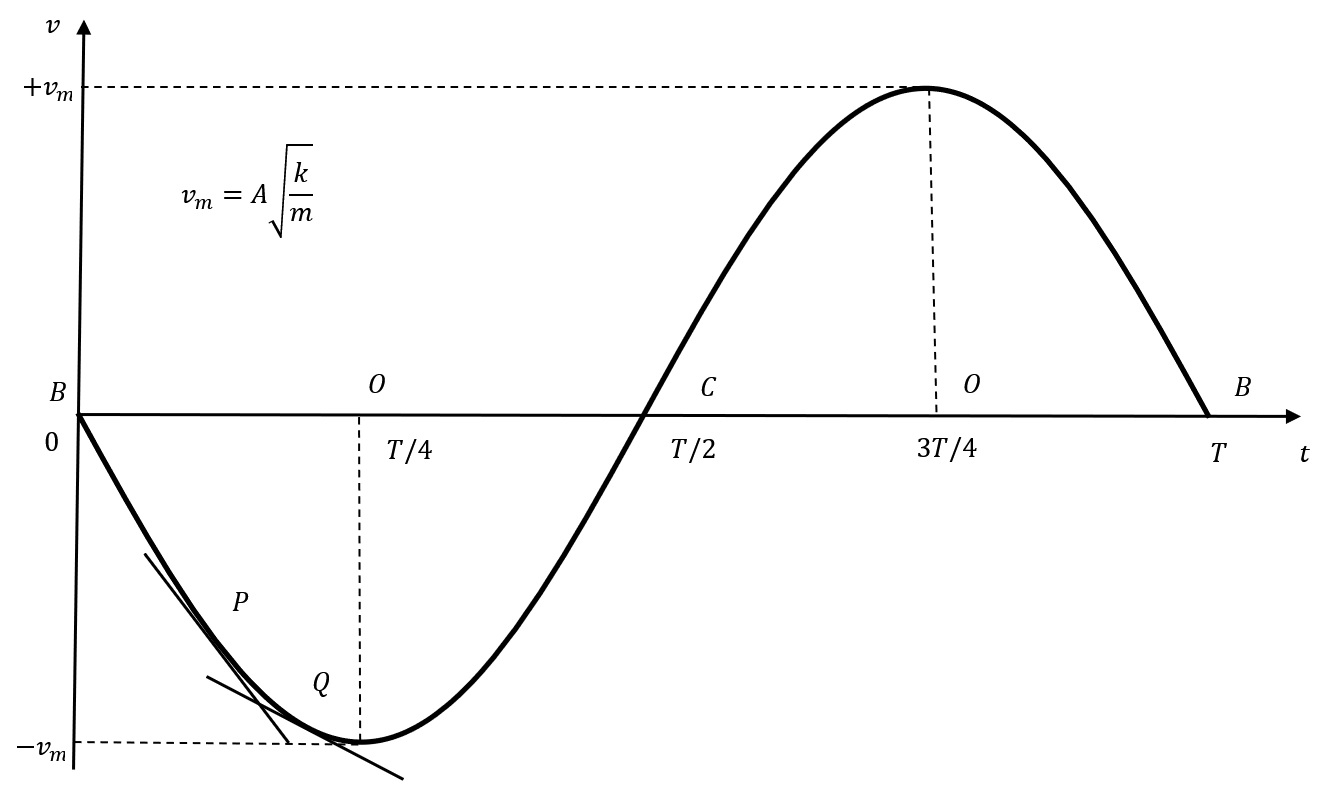}
%\hspace{1cm}。jpg
%\label{velocitygraph}
\caption{Velocity-Time Curve}
\centering
\end{figure}

\subsection{ For $t\in(\fft14\,T, \fft12\,T]$ or $O\rightarrow C$:}
\begin{enumerate}[noitemsep]
	\item At $t=\fft14\,T$ or point $O$, the mass keeps moving to the left; the curve is still below the time axis.
	\item Its speed decreases by Result~6; the curve becomes less and less negative.
	\item The magnitude of acceleration increases by Result~6; the tangent of the curve becomes steeper and steeper.
	\item At $t=\fft12\,T$ or point $C$, the speed is zero.
\end{enumerate}

\subsection{ For $t\in(\fft12\,T, \fft34\,T]$ or $C\rightarrow O$:}
\begin{enumerate}[noitemsep]
	\item At $t=\fft12\,T$ or point $C$, the mass starts moving from rest to the right; the curve is above the time axis.
	\item Its speed increases by Result~7; the curve becomes more and more positive.
	\item The magnitude of acceleration decreases by Result~7; the tangent of the curve becomes more and more shallow.
	\item At $t=\fft34\,T$ or point $O$, the speed achieves its maximum $v_m$ and the acceleration becomes zero; the tangent to the curve is horizontal.
\end{enumerate}

\subsection{ For $t\in(\fft34\,T, T]$ or $O\rightarrow B$:}
\begin{enumerate}[noitemsep]
	\item At $t=\fft34\,T$ or point $O$, the mass keeps moving to the right or in the positive direction; the curve is still above the time axis.
	\item Its speed decreases by Result~8; the curve becomes less and less positive.
	\item The magnitude of acceleration increases by Result~8; the tangent to the curve becomes steeper and steeper.
	\item At $t=T$ or point $B$, the speed is zero.
\end{enumerate}

\section{Conclusion and Discussion}\label{conclusion}
This paper provides a qualitative analysis to SHM without either referring to circular motion or solving differential equation.  
The information contained in the obtained velocity-time curve is rich as far as the rather elementary knowledge, which the analysis is based upon, is concerned. 
In addition, students should always be encouraged to tackle seemingly unreachable new problems with whatever knowledge they have at hand.
The displacement-time curve can similarly be sketched and it is left as an exercise.

%This approach is particularly useful to students without enough mathematical background.
%However, even for students knowing how to solve differential equation, they can still benefit from this approach; qualitative analysis contains intensive physical reasoning that mathematical manipulation cannot provide.

With the experience in such analysis, students will not be surprised, feel comfortable, or even delighted when the sinusoidal solution is obtained from solving differential equation.
It is a good opportunity for them to appreciate the interaction between physics and mathematics.

%One limitation of this qualitative approach is that the exact value of period $T$ cannot be determined. 
%In addition, the velocity-time curve cannot be claimed to be sinusoidal.
%Both have to be determined by other approaches \cite{giancoli, serway}.
 
This approach may be introduced to students right after Newton's second law and Hooke's law are taught. It serves as an example to show how to graphically analyze a motion under varying force instead of a constant one. 
The author is surprised that no textbook employs this approach or at least includes it as an exercise.
Moreover, he sincerely regards this qualitative approach as \textit{the} proper way of teaching SHM to students.

\end{document}